\documentclass[11pt,a4paper]{article}
\usepackage[hyperref]{emnlp2020}
\usepackage{svg}
\usepackage{times}
\usepackage{latexsym}
\usepackage{amsmath}
\usepackage[toc,page]{appendix}
\usepackage{tikz,amsmath,amsthm,graphicx,subfig}
\usepackage{algorithmic,algorithm,varwidth,booktabs}
\usepackage{microtype}

\usepackage{lipsum}
\usepackage{multirow}

\aclfinalcopy % Uncomment this line for the final submission
\title{COVID-19: Comparative Analysis of Methods for Identifying Articles Related to Therapeutics and Vaccines without Using Labeled Data}

\author{Mihir Parmar, Ashwin Karthik Ambalavanan, Hong Guan, Rishab Banerjee,\\\textbf{Jitesh Pabla, Murthy Devarakonda}\\
  Arizona State University \\
  \texttt{mihirparmar@asu.edu, aambalav@asu.edu, hguan6@asu.edu,}\\\texttt{rbanerj8@asu.edu, jpabla1@asu.edu, murthy.devarakonda@asu.edu}}

\date{}

\begin{document}
\maketitle
\begin{abstract}

%Recently, research and development of vaccines and therapeutics for COVID-19 are of global interest. Due to the extensive existing literature (about 59K articles in the CORD-19 dataset), it is challenging to find articles relevant to vaccines and therapeutics among them. This task is even more challenging when we do not have labeled data and do not have time to develop labeled data. In this paper, we applied six different methodologies (i.e., individual scorers) that do not require labeled data, to the task, and conducted a comparative analysis of their performance on the CORD-19 dataset. We also explored the majority voting technique with the methods. Our results show that supervised training using Google search results achieves the best results and that the aggregating the predictions is an open research problem.

Here we proposed an approach to analyze text classification methods based on the presence or absence of task-specific terms (and their synonyms) in the text. We applied this approach to study six different transfer-learning and unsupervised methods for screening articles relevant to COVID-19 vaccines and therapeutics. The analysis revealed that while a BERT model trained on search-engine results generally performed well, it miss-classified relevant abstracts that did not contain task-specific terms. We used this insight to create a more effective unsupervised ensemble.

\end{abstract}

\section{Introduction}
COVID-19 Open Research Dataset (CORD-19) is a machine-readable collection of scientific articles relevant to COVID-19. Finding articles relevant to COVID-19 vaccines and therapeutics in the dataset has many practical uses. Our goal was to study how different transfer-learning and unsupervised methods perform on the task since these methods do not require labeled data.

We formulated the task as a classification problem and considered six different transfer learning or unsupervised techniques for it: (1) BERT's next sentence prediction; (2) Model trained on a different dataset to identify treatments; (2) Clinical semantic text similarity; (4) Lexicon-based semantic similarity; (5) Model trained on Google scholar search results; (6) TF-IDF based scorers. The input to the systems was the title, abstract, and journal name, and the output was one of three labels: vaccine-related, therapeutics-related, or the other.  Where relevant, we used BERT %\cite{bengio2003neural,mikolov2013efficient,mikolov2013distributed,pennington2014glove,devlin2018bert} and its variants.  
\cite{devlin2018bert} and its variants. 

An important challenge was to understand the performance characteristics of these models at a deeper level than the standard performance metrics so that the insight can be used to improve a promising method or develop an effective ensemble. Most error analyses tend to be \textit{ad-hoc}.

We characterized the performance of the methods for four categories that were obtained by taking the cross-product of two factors: (1) whether an article was rated as a positive or negative class; and (2) whether the terms vaccines, therapeutics, and their synonyms appeared in the article (abstract) or not. We manually double-annotated a small (203 articles) test set. Results from our analysis showed that while a Google search-results trained method generally performed well, it miss-classified large portion of articles in one category, and this insight helped us to formulate an effective ensemble, which achieved an F measure of 0.65.

\section{Methods}
\label{sec:methods}
\subsection{Dataset}
We used the COVID-19 Open Research Dataset (CORD-19) \cite{wang2020cord}, which contains around 59K articles previously screened for COVID-19 and associated illnesses and were collected from peer-reviewed publications and archival services. For our experiments, we used the text consisting of [title + abstract + journal name] for each article. We removed the articles that have missing information for one of them, which resulted in about 47K articles. We also manually labeled 203 articles from CORD-19 dataset in three classes: 1) Vaccine, 2) Therapeutics and 3) Other. Each article was judged by two authors of this paper separately. The inter-annotator agreement, measured with kappa score was 0.83. There were 39 vaccine-related articles, 28 therapeutics-related articles, and 136 other articles in the test dataset.

\subsection{Systems}
\subsubsection{NSP-based approach}
\label{sec:nsp}
In this approach, we formulated the task as the Next Sentence Prediction (NSP) in SciBERT \cite{beltagy2019scibert}. We further pre-trained this model using MLM (Masked Learning Model) on the abstract and title text of the CORD-19 dataset. We handcrafted a passage each for vaccines and therapeutics, to use as the first ``sentence". These passages were created from sentences in articles that were positive for vaccine and therapeutics, respectively, and strongly indicated relevancy (based on manual analysis). The passages are shown in Appendix A. 

To test if an article is about therapeutics, the therapeutics passage and the article text (title + abstract + journal name) are used as the “sentence” pair in the next sentence prediction mode of SciBERT. 
If SciBERT predicts that an article is the next sentence with a probability threshold of 0.999 or higher, then this model outputs a “therapeutics” label. Similarly, the article is scored for vaccines, using the vaccine passage as the first sentence. The final predicted label is determined by the higher of the two scores, if the probabilities were greater than 0.999 for both. 

\subsubsection{Clinical Hedges (CH) trained model} 
\label{sec:ch}
In this scorer, we used a SciBERT model that was previously fine-tuned on the Clinical Hedges dataset to identify articles describing treatments. Clinical Hedges \cite{clinical_hedges} is 
a dataset of articles from MEDLINE which were manually annotated for the study-purpose, including therapeutics, etiology, diagnostics and others. Articles were also annotated for other study characteristics that are not of interest here. We defined the task as binary classification, i.e. therapeutics or not therapeutics. 

We combined this model with the NSP model described in Section \ref{sec:nsp} as follows: we used the text documents categorized as therapeutics or vaccine by the NSP model as the input to the Clinical Hedges model at the prediction time. If the Clinical Hedges model categorized the text as therapeutics then the final label is therapeutics
otherwise the label is vaccine. The  Clinical Hedges model consisted of SciBERT (fine-tuned on Clinical Hedges Data) with a Feed-Forward Network (FFN) head to generate the probability distribution for the binary classification.

\subsubsection{STS-based approach}

In this clinical Semantic Textual Similarity (STS) approach, we pre-trained BERT on the Clinical STS dataset from n2c2 Challenge 2019 \cite{cer-etal-2017-semeval} to score semantic similarity between a manually created query and a given text segment. We input the pair ([CLS] + query + [SEP] + text-segment) to the BERT, and used the output CLS representation in a linear regression model to predict similarity in the range of 0 to 5 \cite{cer-etal-2017-semeval}.  Separate queries were manually created for vaccine and therapeutics and are shown in Appendix B. 

The text-segments were obtained by breaking up the concatenation of title, abstract, and the journal name on the sentence boundaries using NLTK. The average of the top $n$ ($n = 3$) text-segment scores was produced as the article-level score \cite{yang2019simple, kotzias2015group}. As suggested in \cite{wang2018medsts}, we used $2.0$ (out of $5.0$) as threshold for classifying an article as positive for a query-type (i.e. therapeutics or vaccine), and the final label was based on the higher of the two scores if both were above 2.0.

\subsubsection{Lexicon-based Similarity Scoring (LSS)}
\label{LSS}
The intuition here is that vaccines-related articles will have many tokens whose embeddings are similar to those of words like ``vaccines". The contextual embeddings for all tokens are first obtained by processing the [title + abstracts + journal name] of articles through the BioBERT model. If a token appear in multiple documents, we take the average of the embeddings of the token from the documents. 

Next, starting with \textit{seed words} -- ``vaccine", ``vaccines", ``vaccination" and ``vaccinations" -- we find additional tokens (\textit{extended-seeds} whose embeddings are within a small cosine distance of the embeddings of the seed words. The cosine distance threshold is set as the max of the closest 1000 pairs. Subsequently, we find the top-$k$ ($k = 50$) words called \textit{representative-words} from the abstract of each article -- the top-$k$ words whose embeddings are closest to any of the extended-seeds. 

Lastly, the \textit{article-score} is the average of the cosine similarity values of the representative-words that appear in its abstract. The process was repeated for the seed words for therapeutics. The final label for an article is the \textit{argmax} of the vaccine and therapeutics article-scores.

%\subsubsection{ElasticBERT}
%
%In this approach, a BERT model (BERT base) is used to generate embedding for a combine text [title + abstract + journal name] for each article in the CORD-19 dataset, which are matched against the embedding of the following queries (for vaccines):
%
%\begin{itemize}
%    \item Vaccine
%    \item Coronavirus vaccine
%\end{itemize}
%
%The top 1000 articles with the highest cosine similarity between their embedding and the query's embedding are labeled as ``vaccine”. The same process is done for ``therapeutics” using the following queries:
%
%\begin{itemize}
%    \item Therapeutics
%    \item Coronavirus therapeutics
%\end{itemize}
%
%The rest of the articles are labeled as ``Other".
%
\subsubsection{Search-results trained approach (GS)}
In this approach, we trained a model using results from the Google search engine, inspired by the knowledge hunting methods \cite{prakash-etal-2019-combining, emami-etal-2018-knowledge}. Google Scholar™ was queried with ``coronavirus vaccine" and ``coronavirus therapeutics", and the respective top 1000 results were scraped. The intersection of the search results for each query and the CORD-19 dataset was designated as the set of positive samples for the corresponding label. 

To obtain negative samples, additional Google Scholar queries that were semantically unrelated to coronavirus vaccines or therapeutics were carefully constructed (see Appendix C). The top 1000 Google Scholar™ results from these queries were intersected with CORD-19 dataset to get the negative samples required. Any articles common between the two positive samples sets and the negative samples set were removed from the negative samples. Any common articles between the vaccine and the therapeutics positive samples were kept in the set for which the article's rank is better. 

A SciBERT model was fine-tuned on 80\% of this weakly labeled dataset, the remaining 20\% was used for validation. Concatenation of article's title, abstract and journal name were given as the input. The trained (fine-tuned) model was used to predict the label for the remaining articles in CORD-19.

\subsubsection{Micro-scorers (MS)}

In this approach, $n$ representative vaccine-related queries ($V_{Q_1}, V_{Q_2}, ..., V_{Q_n}$) and $n$ representative therapeutics-related queries ($T_{Q_1}, T_{Q_2}, ..., T_{Q_n}$) were manually created, and their Cartesian product forms a list of query pairs. The title and abstract were concatenated as the input text for an article. Given an input of the $k^{th}$ article, we first compute its tf-idf vector $a_k$. For each query pair ($V_{Q_i}, T_{Q_j}$), where $i \in [1,n]$ and $j \in [1,n]$, we obtained the tf-idf vectors ($v_i$, $t_j$). The vaccine-score $vs_{k,ij}$, therapeutics-score $ts_{k,ij}$, and other-score $os_{k,ij}$ are computed as follow:
\begin{align*}
    vs_{k,ij} &= \frac{a_k\cdot v_i}{|a_k|\cdot|v_i|}\\
    ts_{k,ij} &= \frac{a_k\cdot t_j}{|a_k|\cdot|t_j|}\\
    os_{k,ij} &= 0.5\cdot(\sqrt{(1-vs_{k,ij})(1-ts_{k,ij})}-\\
         & \sqrt{(1+vs_{k,ij})(1+ts_{k,ij})})
\end{align*}

The intuition behind the other-score $os_{k,ij}$ is that if the angle between $a_k$ and $vs_{k,ij}$ is $\alpha$, the angle between $a_k$ and $ts_{k,ij}$ is $\beta$, then vaccine-score is $\cos\alpha$, therapeutics-score is $\cos\beta$, other-score is $\cos(\frac{\alpha+\beta}{2}+\pi)$. The accumulative scores are computed as follow:

\begin{align*}
    vs_k = \displaystyle\sum_{i,j} vs_{k,ij}, ts_k = \displaystyle\sum_{i,j} ts_{k,ij},
    os_k = \displaystyle\sum_{i,j} os_{k,ij}
\end{align*}
The label corresponding class of the highest score is the final vote for the $k^{th}$ article. In our experiment, we selected four different queries, which are listed in the Appendix B.

\subsubsection{Majority Voting (MV)}

We aggregated the output of all the scorers into one single prediction using majority voting. We did two majority voting scheme: (1) majority of all scorers (MV\_6); (2) majority of three methods (MV\_3) selected based on the performance analysis described in the next section. Each scorer contributes one vote as a label - vaccine, therapeutics, or the other - for each article. The label with the most votes is the overall prediction for the article. Since there are even number of scorers for majority of all, when there is a tie between two labels, a positive class is always preferred and between positive classes one is chosen randomly. 

\subsection{Experiments}

First we used the individual systems described earlier and the majority of all (MV\_6) to predict labels 
(classes) for the manually annotated test dataset. 
%We used standard precision (P), recall (R), and F-measures (F) to evaluate the systems. 
As is customary, we calculated precision (P), recall (R), and F-measures (F) only for the 
positive classes (i.e., Vaccine and Therapeutics).

To characterize performance of the systems compared to the human judgment, 
random 119 of the 204 articles were categorized as shown in Table \ref{tab:scenario}. 
The categories are a cross-product of two factors: 
(1) whether or not specific lexicons are present in the article text (abstract); 
(2) whether or not the article was manually judged as positive class (vaccines or therapeutics).

While 58\% of the articles are in category 4, there were significant number of articles in the other categories as well lending this categorization to interesting analysis of the methods. From the performance of the different systems for these categories, we could draw preliminary conclusions about the scorer design \textit{vis-a-vis} how the subjects of article screening (i.e. vaccines and therapeutics) were discussed in the abstracts. Based on the analysis, a majority of three scorers (MV\_3) was formed and its performance was analyzed on the test dataset. Here is the implementation of all the scorers \footnote{https://github.com/md-labs/covid19-kaggle}.

\begin{table}[!ht]
\centering
\caption{Different categories for analysis. Count and \% out of articles analyzed (119) are shown in parenthesis.}
\label{tab:scenario}
\resizebox{\linewidth}{!}{

\begin{tabular}{cc|c|c|}
\cline{3-4}
&  & \multicolumn{2}{c|}{\begin{tabular}[c]{@{}c@{}}Does the article contain\\words like vaccine, \\therapeutics, or their synonyms?\end{tabular}} \\ 
\cline{3-4} 
&  & YES & NO \\ \hline
\multicolumn{1}{|c|}{\multirow{2}{*}{\begin{tabular}[c]{@{}c@{}}Does the article belong\\to a positive class (vacc. or thera.) \\ according to the manual judgement?\end{tabular}}}                                                     
& YES & \begin{tabular}[c]{@{}c@{}}Category 1\\ (27, 22.69\%)\end{tabular} & \begin{tabular}[c]{@{}c@{}}Category 3\\ (13, 10.92\%)\end{tabular}
\\ \cline{2-4} 
\multicolumn{1}{|c|}{} & NO & \begin{tabular}[c]{@{}c@{}}Category 2\\ (21, 17.65\%)\end{tabular}  & \begin{tabular}[c]{@{}c@{}}Category 4\\ (58, 48.74\%)\end{tabular}   \\ \hline
\end{tabular}
}
\end{table}

\vspace{-0.2cm}
\section{Results}
\label{sec:results}

%Here, we have shown the effectiveness of proposed methodologies in terms of the Precision, Recall, and F-Score analysis. 
Among the individual systems, GS and MS outperformed all the others and achieved the same best F-score (0.50) (See Table \ref{tab:results}). GS achieved the best precision (0.50) while MS achieved the best recall (0.76). NSP and LSS performed rather poorly across the board, and CH and STS achieved moderate performance. As expected, MV\_6 (majority of all) performance was in the middle of the range.  
%In the Table \ref{tab:results}, we highlighted those results with ``Bold". 
%You can also see the comparative performance of all the systems in the form of a bar chart in Figure \ref{fig:results}.

% Table for overall results
\begin{table}[!ht]
\caption{Precision, Recall, and F-measures for different systems. MV\_6 is the majority vote of all, and MV\_3 is majority vote of GS, MS, and CH.}
\label{tab:results}
\resizebox{\linewidth}{!}{
\begin{tabular}{|c||c|c|c|c|c|c||c|c|}
\hline
Systems   & NSP   & CH   & STS   & LSS  & GS            & MS      & MV\_6 & MV\_3           \\ \hline
Precision & 0.12  & 0.42 & 0.32  & 0.21 & \textbf{0.50} & 0.37    & 0.39  & \textbf{0.59}           \\ \hline
Recall    & 0.13  & 0.29 & 0.44  & 0.32 & 0.49          & \textbf{0.76}    & \textbf{0.71}  & 0.69   \\ \hline
F-measure & 0.12  & 0.34 & 0.37  & 0.15 & \textbf{0.50}         & \textbf{0.50}    & 0.49  & \textbf{0.65}   \\ \hline
\end{tabular}
}
\end{table}

%\begin{figure}[!ht]
%    \centering
%    \includegraphics[width=1\linewidth]{images/fscore.pdf}
%    \caption{Graphical representation of the Precision, Recall, and F-Score for different systems.}
%    \label{fig:results}
%\end{figure}

%We also analyzed the performance of all the systems for each category given in Table \ref{tab:scenario}. Here, Table \ref{tab:scenario} represents the description of each category, and the number of total articles falls under that category. We also presented the fraction value to indicate how much a fraction of total articles from the test set each category has. For example, 27 articles fall under category 1, which is 22.69\% of the total test set. 
Table \ref{tab:categoty_results} shows performance of the systems for the four categories introduced in Table \ref{tab:scenario}. We highlighted in bold the best two systems for each category. We note that not all systems performed uniformly across the four categories.
%GS and MS performed better for category 1, correctly classifying 22 and 18 articles respectively out of 27. 
%For category 2, GS and NSP (the overall poorest performing system) performed better, 
%correctly classify 17 and 12 articles out of 21. 
%The overall top performing system, GS, was the worst performer in category 3, 
%while MS and CH performed well for it.
% GS and once again NSP performed well for category 4. 
% It was no surprise that GS performed well on the three categories, but its poor
% performance on category 3 and NSP's second best performance on two categories was unexpected. 
While GS performed well on categories 1, 2, and 4, it did poorly on category 3. On the other hand, MS and CH performed better on category 3. This suggested that combining predictions of GS, CH, and MS might result in a system performing better than any of them. We created majority of three voting system (MV\_3) based on this observation, and found that it achieved the best results on the test dataset (the highest F measure 0.65 and precision 0.59, and the second best recall 0.69).
%Here, we can observe that the GS-based system is performing well for three categories.
\begin{table}[!ht]
\caption{Num. of articles correctly classified by each system out of 119 articles analyzed. (Cat. = Category)}
\label{tab:categoty_results}
\resizebox{\linewidth}{!}{
\begin{tabular}{|c|c||c|c|c|c|c|c||c|c|}
\hline
Cat. & Total & NSP         & CH          & STS & LSS & GS           & MS          & MV\_6 & MV\_3       \\ \hline
1    & { 27} & 03          & 03          & 13  & 08  & \textbf{14}  & \textbf{18} & 16    & \textbf{20} \\ \hline
2    & { 21} & \textbf{12} & 00          & 02  & 01  & \textbf{16}  & 04          & 02    & \textbf{14} \\ \hline
3    & { 13} & 02          & \textbf{09} & 03  & 06  & 01           & \textbf{09} & \textbf{10}    & 08 \\ \hline
4    & { 55} & \textbf{36} & 00          & 18  & 01  & \textbf{49}  & 15          & 13    & \textbf{41} \\ \hline
\end{tabular}
}
\end{table}

\section{Discussion}
\label{sec:discussion}

%In this section, we discuss the advantages and limitations of different proposed systems. Overall performance of the GS-based system is better than the other systems. The reason behind this effective performance is supervised learning using google search. 
Using the labels obtained from the Google Scholar search results to fine-tune the SciBERT model, helped GS perform well across most categories. It, however, performed  poorly on the category 3 articles where terms specific to vaccines and therapeutics were not present. This observation from our analysis suggests the possibility that the term-based relevance matching commonly used in search engines may not be adequate for accurate screening (classification) of articles for a study-topic. 

The robust tf-idf based scoring helped MS to make the better decisions on category 3 articles, in addition to category 1 articles. However, lack of query term discrimination might have undermined its performance for category 2 and 4. Surprisingly better performance of NSP on categories 2 and 4 was due to its tendency to classify most articles as ``Other". CH benefited from transfer learning for category 3. These observations enabled by our analysis method, helped us to combine GS, MS, and CH as a select majority-voting group and achieve the best overall performance.
%Overall, our preliminary results indicate that a combination of GS (search-engine results based weak-supervision) and 
%MS (tf-idf based scorer) or CH (transfer learning of semantic concepts vaccines and therapeutics) 
%is likely to perform well without the task-specific labeled data.

%The overall observation is that there is a need for semantic inference in unsupervised or transfer learning. 

%Here, CH-based system is highly pre-trained on the therapeutics related data, which helps this system to pick up the articles correctly for the positive class even in the absence of the related words. Hence, this system yields better results for category 3. From this, we can see that you need a model that does not only depend on words alone. We believe that it is important to see the words; however, it can not get you all the way in the case of the articles falls under category 3. We need systems that can identify the articles even there are no words present related to the positive classes.

\section{Conclusions}

We proposed a new approach to analyzing performance
of text classification methods. 
We applied this approach to study six transfer learning and unsupervised methods 
for screening articles related to COVID-19 vaccines and therapeutics. 
These methods are of particular value since timely results
can be obtained without the time and effort of generating a large labeled dataset.
We used a novel 2x2 categorization of articles to understand performance
of the systems, and from the analysis formulated an effective voting ensemble of systems.
Our methodology showed that while a weakly supervised model based on search-engine results
performed generally well, it miss-classified articles that 
did not contain task-specific lexicon. Combining it with a tf-idf and a transfer learning 
system yielded better results. The key contribution of this paper 
is the novel approach to analyze text classification methods to gain insights 
into their performance characteristics.

\bibliographystyle{acl_natbib}
\bibliography{emnlp2020}

\begin{thebibliography}{10}
\expandafter\ifx\csname natexlab\endcsname\relax\def\natexlab#1{#1}\fi

\bibitem[{Beltagy et~al.(2019)Beltagy, Lo, and Cohan}]{beltagy2019scibert}
Iz~Beltagy, Kyle Lo, and Arman Cohan. 2019.
\newblock Scibert: A pretrained language model for scientific text.
\newblock In \emph{Proceedings of the 2019 Conference on Empirical Methods in
  Natural Language Processing and the 9th International Joint Conference on
  Natural Language Processing (EMNLP-IJCNLP)}, pages 3606--3611.

\bibitem[{Cer et~al.(2017)Cer, Diab, Agirre, Lopez-Gazpio, and
  Specia}]{cer-etal-2017-semeval}
Daniel Cer, Mona Diab, Eneko Agirre, I{\~n}igo Lopez-Gazpio, and Lucia Specia.
  2017.
\newblock {S}em{E}val-2017 task 1: Semantic textual similarity multilingual and
  crosslingual focused evaluation.
\newblock In \emph{Proceedings of the $11^{th}$ International Workshop on
  Semantic Evaluation ({S}em{E}val-2017)}, pages 1--14, Vancouver, Canada.
  Association for Computational Linguistics (ACL).

\bibitem[{Devlin et~al.(2018)Devlin, Chang, Lee, and
  Toutanova}]{devlin2018bert}
Jacob Devlin, Ming-Wei Chang, Kenton Lee, and Kristina Toutanova. 2018.
\newblock Bert: Pre-training of deep bidirectional transformers for language
  understanding.
\newblock \emph{arXiv preprint arXiv:1810.04805}.
\newblock \{Last Accessed: May 24, 2019\}.

\bibitem[{Emami et~al.(2018)Emami, De~La~Cruz, Trischler, Suleman, and
  Cheung}]{emami-etal-2018-knowledge}
Ali Emami, Noelia De~La~Cruz, Adam Trischler, Kaheer Suleman, and Jackie
  Chi~Kit Cheung. 2018.
\newblock A knowledge hunting framework for common sense reasoning.
\newblock In \emph{Proceedings of the 2018 Conference on Empirical Methods in
  Natural Language Processing}, pages 1949--1958, Brussels, Belgium.
  Association for Computational Linguistics.

\bibitem[{Kotzias et~al.(2015)Kotzias, Denil, De~Freitas, and
  Smyth}]{kotzias2015group}
Dimitrios Kotzias, Misha Denil, Nando De~Freitas, and Padhraic Smyth. 2015.
\newblock From group to individual labels using deep features.
\newblock In \emph{Proceedings of the 21th ACM SIGKDD International Conference
  on Knowledge Discovery and Data Mining}, pages 597--606.

\bibitem[{Prakash et~al.(2019)Prakash, Sharma, Mitra, and
  Baral}]{prakash-etal-2019-combining}
Ashok Prakash, Arpit Sharma, Arindam Mitra, and Chitta Baral. 2019.
\newblock Combining knowledge hunting and neural language models to solve the
  {W}inograd schema challenge.
\newblock In \emph{Proceedings of the $57^{th}$ Annual Meeting of the
  Association for Computational Linguistics}, pages 6110--6119, Florence,
  Italy. Association for Computational Linguistics.

\bibitem[{Wang et~al.(2020)Wang, Lo, Chandrasekhar, Reas, Yang, Eide, Funk,
  Kinney, Liu, Merrill et~al.}]{wang2020cord}
Lucy~Lu Wang, Kyle Lo, Yoganand Chandrasekhar, Russell Reas, Jiangjiang Yang,
  Darrin Eide, Kathryn Funk, Rodney Kinney, Ziyang Liu, William Merrill, et~al.
  2020.
\newblock \href
  {https://www.kaggle.com/allen-institute-for-ai/CORD-19-research-challenge}
  {Cord-19: The covid-19 open research dataset}.
\newblock \emph{arXiv preprint arXiv:2004.10706}.
\newblock \{Last Accessed: Apr 25, 2020\}.

\bibitem[{Wang et~al.(2018)Wang, Afzal, Fu, Wang, Shen, Rastegar-Mojarad, and
  Liu}]{wang2018medsts}
Yanshan Wang, Naveed Afzal, Sunyang Fu, Liwei Wang, Feichen Shen, Majid
  Rastegar-Mojarad, and Hongfang Liu. 2018.
\newblock Medsts: a resource for clinical semantic textual similarity.
\newblock \emph{Language Resources and Evaluation}, pages 1--16.

\bibitem[{Wilczynski~NL(2005)}]{clinical_hedges}
Haynes RB Hedges~Team Wilczynski~NL, Morgan~D. 2005.
\newblock An overview of the design and methods for retrieving high-quality
  studies for clinical care.
\newblock \emph{BMC Med Inform Decis Mak.}, (Jun):5--20.

\bibitem[{Yang et~al.(2019)Yang, Zhang, and Lin}]{yang2019simple}
Wei Yang, Haotian Zhang, and Jimmy Lin. 2019.
\newblock Simple applications of bert for ad hoc document retrieval.
\newblock \emph{arXiv preprint arXiv:1903.10972}.

\end{thebibliography}

\begin{appendices}

\section{Queries used in NSP-based Approach}
\label{query_nsp}
Vaccine-related query:

Vaccine is a substance used to stimulate the production of antibodies and provide immunity against diseases. They are treated to act as an antigen without inducing the disease. When the virulent version of an agent comes along, the immune system is prepared to respond due to the generation of B cells (memory and plasma cells), which will generate antibodies that will bind to pathogens and destroy them. Vaccine researchers are working on the development of a vaccine candidate expressing the viral spike protein of SARS-CoV-2 using a messenger RNA vaccine. Scientists are also focusing on the development of a chimpanzee adenovirus-vectored vaccine candidate against COVID-19. In addition, scientists are also working to see if vaccines developed for SARS coronavirus are effective against COVID-19.\\\newline
Therapeutics-related query:

Therapeutics is the branch of medicine concerned with the therapeutics of disease and the action of remedial agents. There is no specific antiviral therapy and therapeutics given by doctors is largely supportive, consisting of supplemental oxygen and conservative fluid administration. Drugs like Chloroquine, Hydroxychloroquine, Lopinavir, Ritonavir, Azithromycin and Tocilizumab are being prescribed by doctors in ICU testing. The drug Remdesivir has shown promise against other coronaviruses in animal models. Patients with respiratory failure require intubation. Patients in shock require urgent fluid resuscitation and administration of empiric antimicrobial therapy. Corticosteroid therapy is not recommended for viral pneumonia; however, use may be considered for patients with refractory shock or acute respiratory distress syndrome.

\section{Queries used in Micro-scorers (MS) and STS-based Approach}
\label{query_sts}
Vaccine-related query:
\begin{itemize}
\item vaccine vaccination dose antitoxin serum immunization inoculation for COVID-19 or coronavirus related research work.
% \item Methods evaluating potential complication of Antibody-Dependent Enhancement (ADE) in vaccine recipients of COVID-19 or coronavirus.
% \item Methods evaluating potential complication of Antibody-Dependent Enhancement (ADE) in vaccine recipients.
% \item Exploration of use of best animal models and their predictive value for a human vaccine.
\end{itemize}
Therapeutics-related query:
\begin{itemize}
\item Therapeutics therapeutics therapy drug antidotes cures remedies medication prophylactic restorative panacea for COVID-19 or coronavirus related research work.
% \item Effectiveness of drugs like naproxen, clarithromycin, and minocyclinethat being developed that may exert effects on viral replication and tried to treat COVID-19 patients.
% \item Clinical and bench trials to investigate less common viral inhibitors against COVID-19 such as naproxen, clarithromycin, and minocyclinethat that may exert effects on viral replication.
% \item Effectiveness of drugs being developed and tried to treat COVID-19 patients.
\end{itemize}

\section{Queries used in Search-results trained approach (GS)}
\label{query_google}
\begin{itemize}
    \item Coronavirus transmission, incubation and environment stability
    \item Coronavirus ethical and social science considerations
    \item Coronavirus information sharing and inter-sectoral collaboration
\end{itemize}

\end{appendices}

\end{document}